\documentclass[aps,prb,twocolumn,showpacs,groupedaddress]{revtex4-1}
\usepackage{amssymb}

\usepackage{graphicx}

\begin{document}

\title[]{Giant Magnetoelectric Effect in HoAl$_3$(BO$_3$)$_4$}

\author{K.-C. Liang$^1$, R. P. Chaudhury$^1$, B. Lorenz$^1$, Y. Y. Sun$^1$, L. N. Bezmaternykh$^2$, V. L. Temerov$^2$, and C. W. Chu$^{1,3}$}

\affiliation{$^1$ TCSUH and Department of Physics, University of Houston, Houston, Texas 77204-5002, USA}

\affiliation{$^2$ Institute of Physics, Siberian Division, Russian Academy of Sciences, Krasnoyarsk, 660036 Russia}

\affiliation{$^3$ Lawrence Berkeley National Laboratory, 1 Cyclotron Road, Berkeley, California 94720, USA}

\begin{abstract}
A giant magnetoelectric polarization is found in HoAl$_3$(BO$_3$)$_4$. The polarization in transverse field geometry at 70 kOe reaches 3600 $\mu C/m^2$ which is significantly higher than reported values of linear magnetoelectric or even multiferroic compounds. The magnetostrictive effect is also measured and compared with the magnetoelectricity. The results show that spin-lattice coupling in HoAl$_3$(BO$_3$)$_4$ is extremely strong and that the magnetic field causes a large polar distortion of the ionic positions in the unit cell.
\end{abstract}


\maketitle

The magnetoelectric effect, reflecting the coupling between magnetic (electric) fields and electric (magnetic) polarizations in matter, has inspired researchers for more than a century. The experimental discovery of the linear magnetoelectric coupling in antiferromagnetic Cr$_2$O$_3$ by Rado and Folen\cite{rado:61} in 1961 has started a new field of research and many magnetoelectric materials have been identified and investigated in the ensuing years.\cite{schmid:03} Materials with a large magnetoelectric effect are of interest for the development of new technologies and applications as magnetoelectric sensors, memory elements, etc.

The magnetoelectric coupling can be derived from the expansion of the free energy with respect to magnetic (H) and electric (E) fields. The linear coupling term involving electric and magnetic fields, $\alpha_{ij}E_iH_j$, defines the magnetoelectric susceptibility tensor $\alpha_{ij}$. This term gives rise to a linear magnetic-field induced contribution to the electric polarization, $P_i=\alpha_{ij}H_j$.\cite{schmid:03} Unfortunately, the linear magnetoelectric effect is frequently forbidden by symmetry or relatively small.\cite{brown:68,fiebig:05} In fact, the magnetoelectric coefficients reported for a variety of materials do not exceed 100 ps/m (see table 3 in Ref. \cite{schmid:03}). For comparison, the magnitude of $\alpha_{33}$ of Cr$_2$O$_3$, the first magnetoelectric material reported, is less than 5 ps/m.\cite{rado:61} Only TbPO$_4$ was found with a significantly larger coefficient of 730 ps/m at a temperature of 1.5 K,\cite{rivera:09} corresponding to a field-induced polarization of 58 $\mu$C/m$^2$ at 1 kOe. Besides the linear magnetoelectric effect the bilinear coupling also exists and may result in a sizable magnetoelectric polarization, $P\sim H^2$, especially at higher fields.

In the search for materials with larger magnetoelectric couplings the recent attention has shifted toward the study of multiferroic compounds\cite{fiebig:05,eerenstein:06,khomskii:06,cheong:07,tokura:07} since the coexistence of magnetic and ferroelectric orders in this class of materials is commonly associated with large values of electric and magnetic susceptibilities in some temperature range, allowing for larger linear magnetoelectric coefficients. Relatively large values of the ferroelectric polarization have been reported in multiferroics like TbMnO$_3$ (800 $\mu$C/m$^2$),\cite{kimura:03} GdMn$_2$O$_5$ (1200 $\mu$C/m$^2$),\cite{inomata:96} and DyMnO$_3$ (2500 $\mu$C/m$^2$).\cite{kimura:05} The tunability of the ferroelectric polarization in magnetic fields has been shown for some multiferroics mainly through the coupling of the field to the magnetic order and field-induced magnetic phase transitions.\cite{higashiyama:04,hur:04,taniguchi:06}

Another class of noncentrosymmetric compounds, rare earth iron borates - RFe$_3$(BO$_3$)$_4$ (R=rare earth, Y), have been recently shown to exhibit a significant magnetoelectric effect.\cite{zvezdin:05,krotov:06,zvezdin:06,yen:06,zvezdin:06b,chaudhury:09} The RFe$_3$(BO$_3$)$_4$ system belongs to the trigonal system with space group R32 \cite{joubert:68} and they show a wealth of magnetic phase transitions with the ordering of Fe-spins as well as rare earth moments. The magnetoelectric properties and magnetic phase diagram of these compounds are very complex because of the magnetic exchange of the Fe-spins giving rise to antiferromagnetic (AFM) order and their coupling to the rare-earth f-moments with a strong magnetic anisotropy.\cite{kadomtseva:10} We have shown recently that the transition metal (Fe) is not necessary to establish large magnetoelectricity in this class of materials and that the isostructural compound TmAl$_3$(BO$_3$)$_4$ displays a sizable bilinear magnetoelectric effect \cite{chaudhury:10b}. These results led us to further studies of the RAl$_3$(BO$_3$)$_4$ system in the search for materials with larger magnetoelectric polarizations.

In this communication we report the discovery of a giant magnetoelectric effect in HoAl$_3$(BO$_3$)$_4$ that exceeds the typical field-induced polarization changes of magnetoelectric and multiferroic materials at high magnetic fields significantly.

\begin{figure}
\begin{center}
\includegraphics[angle=0, width=2.5 in]{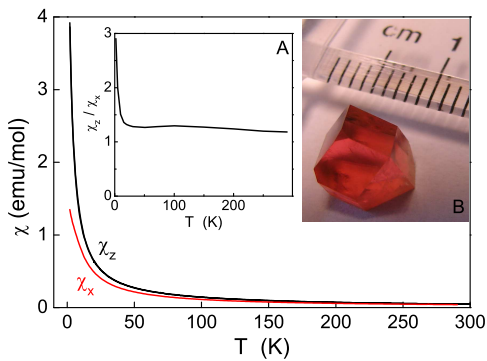}
\end{center}
\caption{(Color online) Magnetic susceptibility of HoAl$_3$(BO$_3$)$_4$ with the field oriented along the $x$ and $z$ axes. Inset A: Magnetic anisotropy, $\chi_z$/$\chi_x$. Inset B: Photograph of a single crystal. The small equilateral triangle on top of the crystal defines the hexagonal symmetry with $x$ along the edge and $z$ perpendicular to the triangle.}
\end{figure}

Single crystals of HoAl$_3$(BO$_3$)$_4$ have been grown from the flux as described earlier.\cite{temerov:08} The crystals show smooth facets and their color varies between yellow and pink (Inset B in Fig. 1). The crystals were cut and polished according to the demands of various experiments. The orientation of the samples was determined by Laue X-ray diffraction. The magnetic susceptibility was measured in a 50 kOe magnetometer (Quantum Design) along different field orientations. For the remaining discussions we will use an orthogonal system ($x$,$y$,$z$) with $x$ and $z$ along the hexagonal $a$- and $c$-axes and $y$ perpendicular to $x$ and $z$. Magnetoelectric measurements were conducted by measuring the current between two contacts (made from silver paint) attached to a parallel plate sample while sweeping the magnetic field. Temperature and field control were provided by the Physical Property Measurement System (Quantum Design). The field-induced polarization was determined from the charge obtained by integrating the magnetoelectric current with respect to time. Magnetostriction measurements were performed using the strain gage method.\cite{chaudhury:10b} The magnetoelectric polarization and the magnetostriction were both measure in longitudinal and transverse field orientations.

The magnetic susceptibilities $\chi_x$ and $\chi_z$, shown in Fig. 1, reveal that HoAl$_3$(BO$_3$)$_4$ is a nearly isotropic paramagnet. $\chi_y$ (not shown in Fig. 1) was found to be identical to $\chi_x$ within the experimental error limits. The anisotropy ratio, $\chi_z$/$\chi_x$, is approximately 1.2 over a large temperature range, increasing to about 3 at the lowest temperatures (Inset A). The nearly isotropic magnetism distinguishes HoAl$_3$(BO$_3$)$_4$ from other RAl$_3$(BO$_3$)$_4$ which either exhibit a significantly stronger easy plane (R = Tm,\cite{chaudhury:10b} Er) or easy axis anisotropy (R = Tb). The inverse susceptibility scales linear with temperature and the extracted effective magnetic moment of 10.6 $\mu_B$ is in perfect agreement with the expected value for the f-moment of the Ho$^{3+}$ ion.

\begin{figure}
\begin{center}
\includegraphics[angle=0, width=2.5 in]{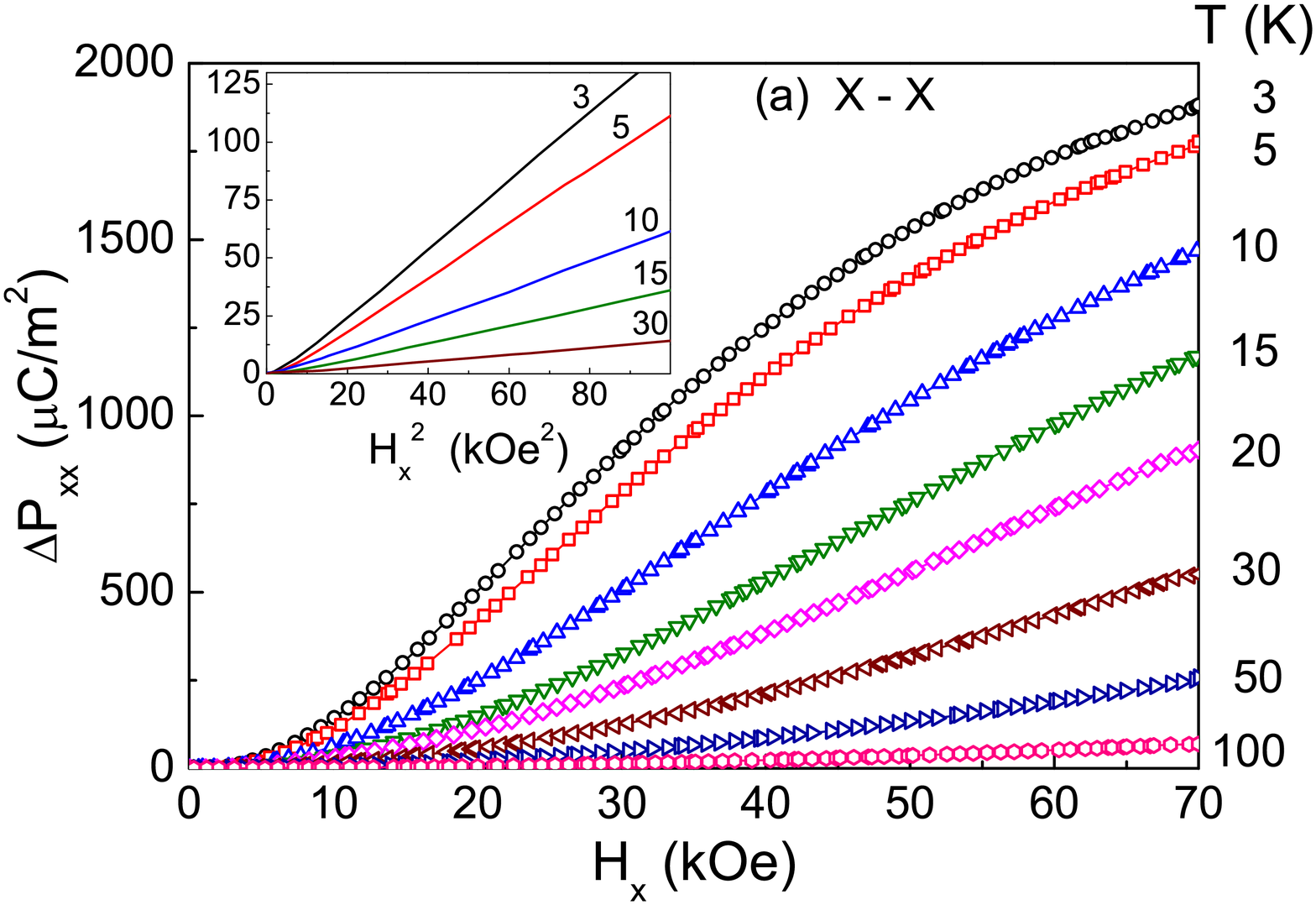}
\includegraphics[angle=0, width=2.5 in]{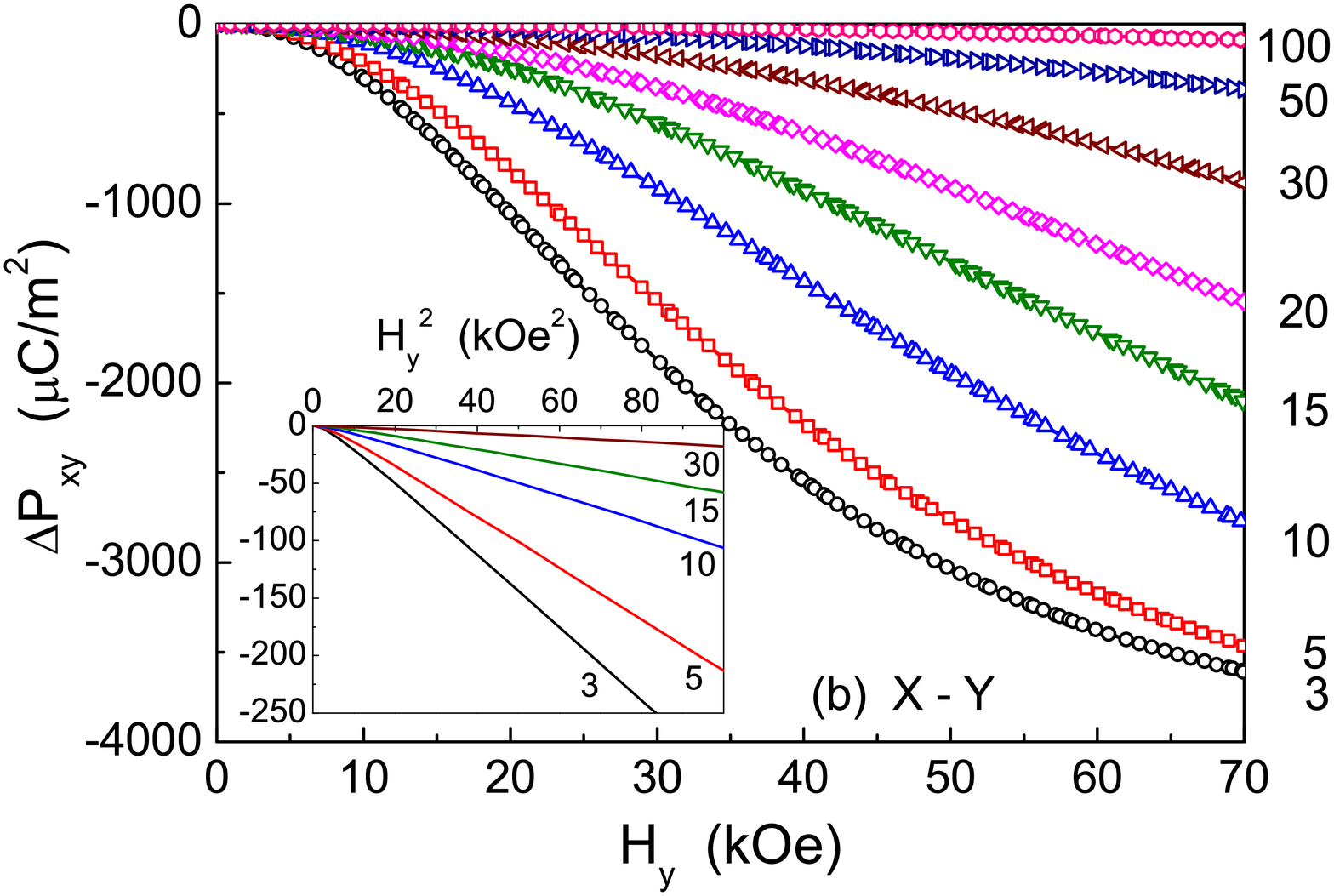}
\includegraphics[angle=0, width=2.5 in]{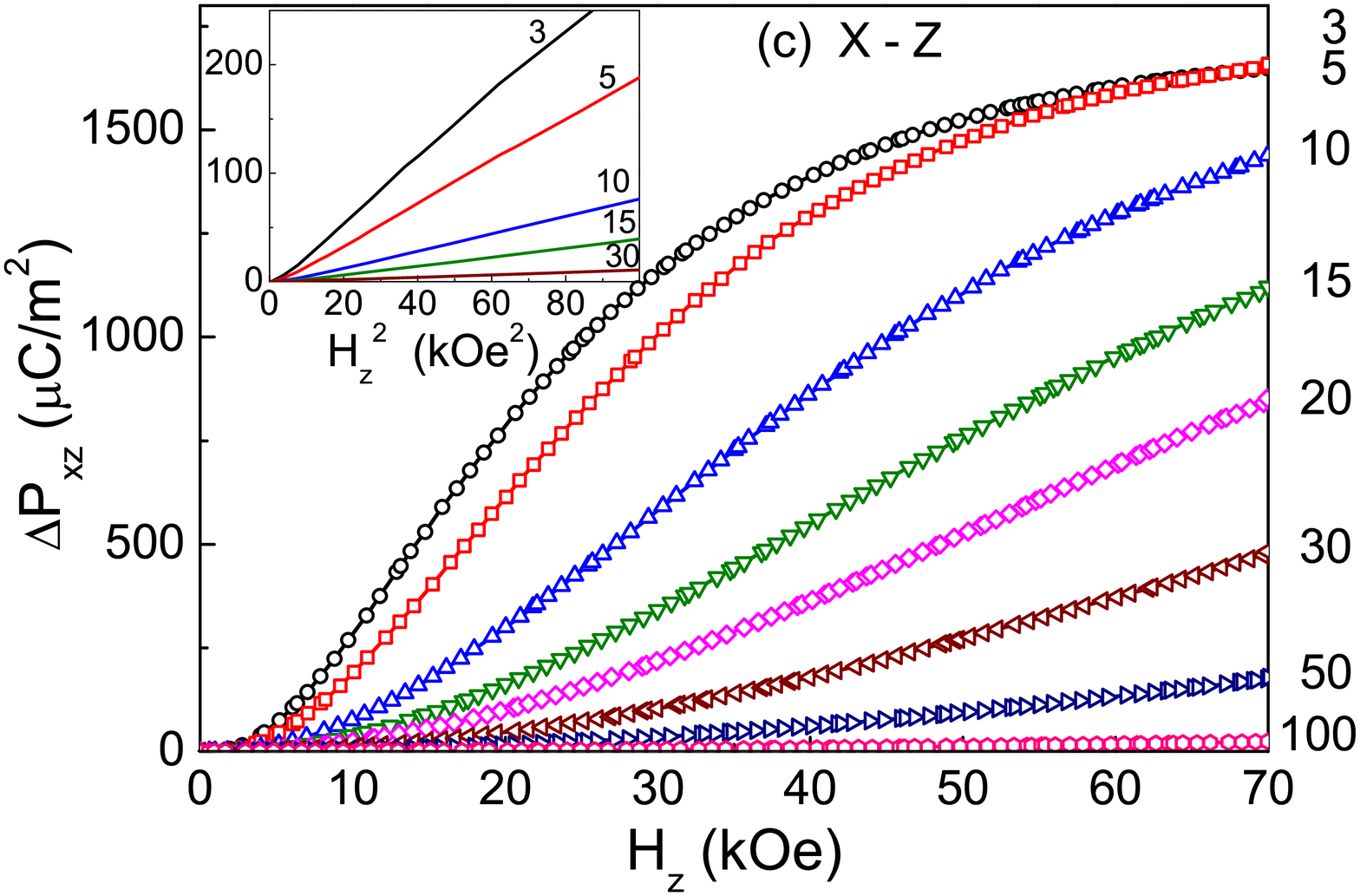}
\end{center}
\caption{(Color online) Magnetoelectric polarization of HoAl$_3$(BO$_3$)$_4$ in (a) longitudinal and (b), (c) transverse field orientations. The labels next to the curves indicate the temperature of the measurements. Insets show the $P\propto H^2$ scaling in magnetic fields below 10 kOe.}
\end{figure}

The magnetoelectric polarizations along the $x$ axis in fields along $x$ ($\Delta P_{xx}$), $y$ ($\Delta P_{xy}$), and $z$, ($\Delta P_{xz}$), are shown in Figures 2a, 2b, and 2c, respectively. At low field the polarizations are proportional to $H^2$ indicating that the magnetoelectric effect is bilinear up to 10 kOe, as shown in the insets of Fig. 2. The polarization reaches very large values at higher magnetic fields. It is important to maintain the same contact arrangement and electrical wire connections in all measurements of the longitudinal and transverse magnetoelectric polarizations. While the sign of $\Delta P_{ij}$ is determined by the original orientation of the non-centrosymmetric crystal, the relative sign of $\Delta P_{ij}$ for different field orientations is intrinsic and provides additional insights. The longitudinal polarization, $\Delta P_{xx}$, shown in Fig. 2a, rises to 1900 $\mu C/m^2$ at 3 K and 70 kOe. This value is almost an order of magnitude larger than the polarization of TmAl$_3$(BO$_3$)$_4$ reported earlier.\cite{chaudhury:10b}

\begin{figure}
\begin{center}
\includegraphics[angle=0, width=2.5 in]{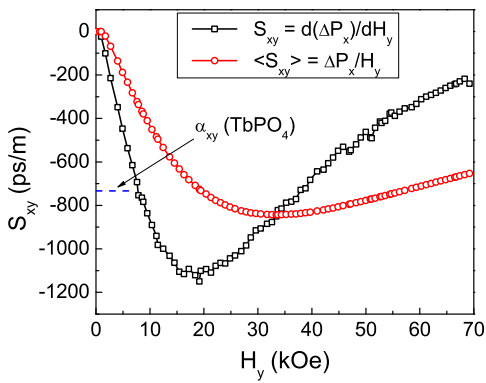}
\end{center}
\caption{(Color online) Open squares: Transverse magnetoelectric sensitivity, $S_{xy}=d(\Delta P_{x})/dH_y$, of HoAl$_3$(BO$_3$)$_4$. Open circles: Average sensitivity $\Delta P_x/H_y$. The dashed line shows the sensitivity of TbPO$_4$ (Note that above 8 kOe the magnetoelectricity in TbPO$_4$ breaks down).}
\end{figure}

The transverse polarization, $\Delta P_{xy}$, turns out to be even larger in magnitude but with opposite sign (Fig. 2b). The maximum polarization at 70 kOe is -3600 $\mu C/m^2$ which exceeds the values of all other magnetoelectric compounds at this field. It is noteworthy that the magnetoelectric polarization values reported here are more than 10 times larger than those of NdFe$_3$(BO$_3$)$_4$ at 70 kOe, the compound with the largest magnetoelectric effect in the R-Fe borate system.\cite{zvezdin:06b} The transverse magnetoelectric polarization $\Delta P_{xz}$, shown in Fig. 2c, also rises to relatively large values of up to 1650 $\mu C/m^2$ and its sign is positive, similar to $\Delta P_{xx}$.

We have also studied the polarization along the $z$-axis in longitudinal and transverse magnetic fields. The magnetoelectric polarization found is two orders of magnitude smaller than the $x$-axis values. The largest $\Delta P_z$ measured at 3 K was only 35 $\mu C/m^2$ in a transverse field of 70 kOe.

The magnetoelectric sensitivity, $S_{ij}=d(\Delta P_i)/dH_j$, as a measure of the polarization change induced by a change of magnetic field, is a relevant quantity for technological utilizations of the magnetoelectric effect. For linear magnetoelectrics $S_{ij}$ is a constant identical to $\alpha_{ij}$. $S_{xy}$ is plotted for the transverse configuration in Fig. 3 (open squares). Since $\Delta P_{xy}$ is a nonlinear function of H$_y$, the sensitivity changes with the field rising to a maximum of $\mid S_{xy}\mid$=1130 ps/m at about 20 kOe. The ratio $\langle S_{xy}\rangle=\Delta P_x/H_y$ (open circles in Fig. 3) describes the average rate of change of the polarization upon increasing the field from 0 to H. It still rises in magnitude to 850 ps/m, above the value for the linear transverse coefficient of TbPO$_4$ (horizontal dashed line in Fig. 3).

\begin{figure}
\begin{center}
\includegraphics[angle=0, width=2.5 in]{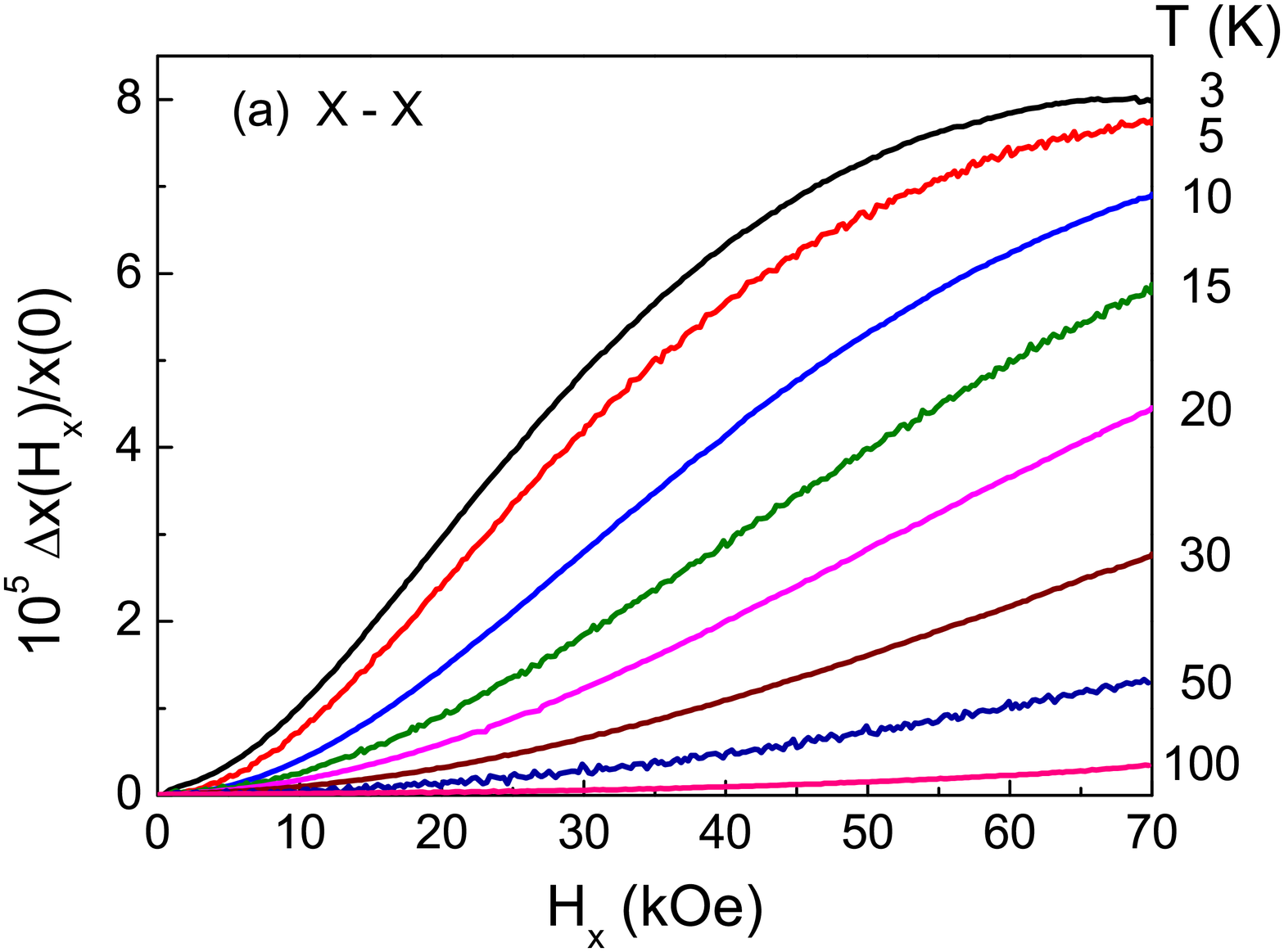}
\includegraphics[angle=0, width=2.5 in]{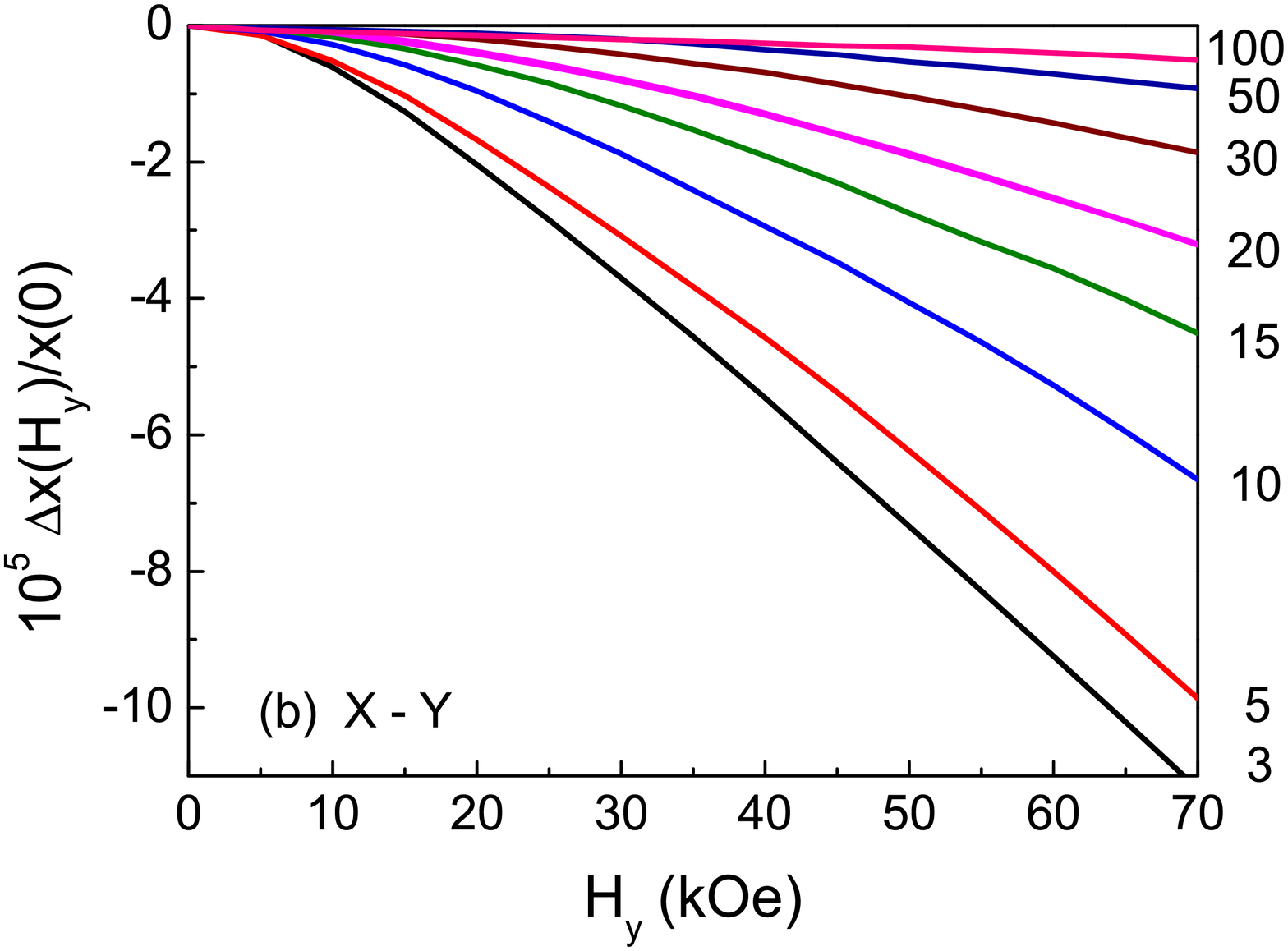}
\includegraphics[angle=0, width=2.5 in]{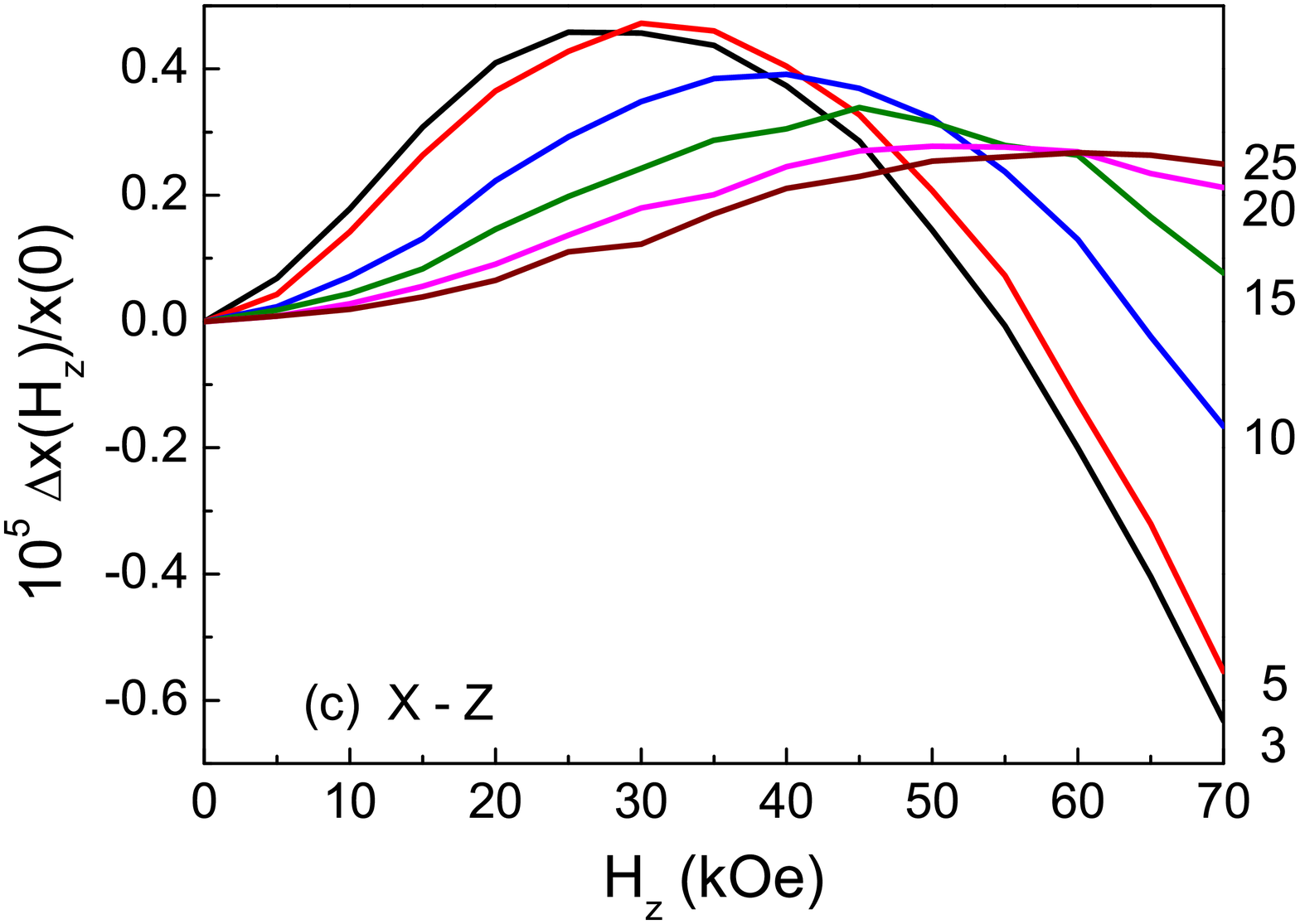}
\end{center}
\caption{(Color online) $a$-axis magnetostriction of HoAl$_3$(BO$_3$)$_4$ in (a) longitudinal and (b), (c) transverse field orientations. The labels next to the curves indicate the temperature of the measurements.}
\end{figure}

It is worth noting that the relative sign of $\Delta P_{ij}$ in Fig. 2 depends on the magnetic field orientation, where $\Delta P_{xy}$ has the opposite sign to that of $\Delta P_{xx}$ and $\Delta P_{xz}$. The sign and magnitude of $\Delta P_{ij}$ are associated with the microscopic structural changes caused by the external field. While the zero-field structure is non-polar (space group R32), the magnetic field apparently changes the symmetry to allow for a macroscopic polarization. The microscopic distortions also result in a change of the lattice parameters and a macroscopic length change (magnetostriction). The results of longitudinal and transverse magnetostriction measurements are shown in Fig. 4 for the same directions of measurements and applied fields as in the polarization measurements of Fig. 2.

The $x$-axis in longitudinal field (X-X configuration) expands with H (Fig. 4a). Although the magnitude of $\Delta L(H)/L(0)$ is comparable with the magnetostriction in TmAl$_3$(BO$_3$)$_4$,\cite{chaudhury:10b} the opposite sign (expansion instead of compression) indicates significant differences between these two isostructural compounds. Fields directed along the $y$-axis (X-Y geometry) result in a large compression of the $x$-axis (Fig. 4b) in HoAl$_3$(BO$_3$)$_4$. In contrast, The $z$-axis field causes a relatively minor change of the $x$-axis (X-Z geometry in Fig. 4c). Interestingly, the length along $x$ first increases in small $z$-axis fields and then decreases at higher H$_z$. The overall change in this case is 20 times smaller than in the X-Y configuration (Fig. 4b).

Comparing the field-induced polarization (Fig. 2) and the magnetostriction (Fig. 4), it becomes obvious that a simple correlation between the polarization and length change along the $x$-axis does not exist for all three field orientations. While the field dependencies of $\Delta P_{xx}$ and $\Delta x(H_x)/x(0)$ are similar (Figs. 2a and 4a) and both quantities scale well with one another, an analogous scaling behavior is not obvious for $\Delta P_{xy}$ and $\Delta x(H_y)/x(0)$ as well as $\Delta P_{xz}$ and $\Delta x(H_z)/x(0)$ (Figs. 2b,c and 4b,c). It is therefore concluded that the large values of $\Delta P_{ij}$ arise from complex atomic displacements within the unit cell and that these displacements strongly depend on the orientation of the external field. High-resolution scattering (X-ray, neutron) experiments could reveal the nature of the distortions and the associated change of symmetry but it is not clear whether the resolution of those techniques would be sufficient. Further studies should target the field-induced structural changes resulting in the large magnetoelectric effect on a microscopic level.

The magnetoelectric response of HoAl$_3$(BO$_3$)$_4$ is strong in the high-field range ($>$10 kOe) where the magnetoelectricity of linear magnetoelectrics usually breaks down (above 8 kOe in TbPO$_4$)\cite{kahle:86,mensinger:93} or decreases significantly (above 5 kOe in hexaferrites).\cite{kitagawa:10} It also allows for a continuous change of the polarization with magnetic fields in contrast to the magnetoelectric response of many multiferroic compounds, where the largest polarization changes happen in a narrow field or temperature range due to a phase transition in the magnetic system.

By reviewing the magnetoelectricity in the RAl$_3$(BO$_3$)$_4$ system, we found that the effect is closely related to the magnetic anisotropy of the compound and decreases as the latter increases. For example, the field-induced polarization increases from zero in TbAl$_3$(BO$_3$)$_4$ (with a strong uniaxial $z$-axis anisotropy\cite{chaudhury:11b}) through ErAl$_3$(BO$_3$)$_4$ (strongest easy plane anisotropy\cite{chaudhury:11b}) and TmAl$_3$(BO$_3$)$_4$\cite{chaudhury:10b} to HoAl$_3$(BO$_3$)$_4$ (nearly isotropic) as their magnetic anisotropy decreases. This suggests that a large magnetoelectricity is realized most favorably in the magnetically isotropic members of the RAl$_3$(BO$_3$)$_4$ compounds.

In summary, we have discovered a giant magnetoelectric effect in HoAl$_3$(BO$_3$)$_4$ of 3600 $\mu C/m^2$ along the $x$-axis when 70 kOe is applied along the $y$-axis. The magnitude of this effect at higher fields exceeds the field-induced polarization values of linear magnetoelectric materials including the current record holder TbPO$_4$, as well as the values found in the isostructural system RFe$_3$(BO$_3$)$_4$. We also found that the magnetoelectric effect of the RAl$_3$(BO$_3$)$_4$ system increases with decreasing magnetic anisotropy. The microscopic picture is yet to be revealed. From the complimentary magnetostriction measurements it is concluded that the magnetic field couples strongly to the ions on a microscopic scale and that the origin of the large magnetoelectric effect lies in the relative ionic displacements in the unit cell of HoAl$_3$(BO$_3$)$_4$ accompanied by a field-induced symmetry change to a polar structure. The current results suggest to search for other compounds of similar structure but with alterations of the rare earth or the d/p electron elements.

\begin{acknowledgments}
This work is supported by the DoE, the US AFOSR, the
T.L.L. Temple Foundation, the J. J. and R. Moores Endowment, and the
State of Texas through the TCSUH.
\end{acknowledgments}


%

\end{document}